\documentclass{aa}

\usepackage{graphicx}
\usepackage{color}
\begin{document}
\title{GeV-TeV gamma-ray light curves expected in the IC $e^\pm$ pair cascade
model for massive binaries: Application to LS 5039}

   \author{W. Bednarek}

   \offprints{W. Bednarek}

   \institute{Department of Experimental Physics, University of \L \'od\'z,
              90-236 \L \'od\'z, ul. Pomorska 149/153, Poland}

   \date{Received ...., .... ; accepted ...., ....}

 \abstract
    {
   TeV gamma-ray emission from  two massive binaries of the microquasar type, LS 5039 and LS I +61$^{\rm o}$ 303, show clear variability with their orbital periods.
}
{
Our purpose is to calculate the GeV and TeV $\gamma$-ray light curves from the massive binary LS 5039 which are expected in the specific Inverse Compton $e^\pm$ pair cascade model. This model successfully predicted the basic features of the high energy $\gamma$-ray emission from LS 5039 and LS I +61$^{\rm o}$ 303. 
}
{
   In the calculations we apply the Monte Carlo code which follows the IC $e^\pm$ pair cascade in the anisotropic radiation of the massive star.    }
  {
   The $\gamma$-ray light curves and spectra are obtained for different parameters of the acceleration scenario and the inclination angles of the binary system.  
It is found that the GeV and TeV $\gamma$-ray light curves should be anti-correlated. 
This feature can be tested in the near future by the simultaneous observations of LS 5039 
with the AGILE and GLAST telescopes in GeV energies and the Cherenkov telescopes in the TeV 
energies. 
Considered model also predicts a broad maximum in the TeV $\gamma$-ray light curve between 
the phases $\sim 0.4-0.8$ consistently with the observations of LS 5039 by the HESS telescopes. 
Moreover, we predict additional dip in the TeV light curve for large inclination angles $\sim 60^{\rm o}$.
This feature could serve as a diagnostic for independent measuring of the inclination angle of
this binary system indicating also on the presence of a neutron star in LS 5039.}
   {}


\titlerunning{Gamma-ray light curves from LS 5039}

   \maketitle

%
%
%
\section{Introduction}

Recently, two massive binary systems of the microquasar type have been discovered
as TeV $\gamma$-ray sources, LS 5039 (Aharonian et al.~2005) and LS I +61$^{\rm o}$ 303 
(Albert et al.~2006)). The TeV $\gamma$-ray emission is modulated with the
orbital periods of these two binary systems (Albert et al.~2006, Aharonian et al.~2006a). 
This feature strongly indicate that high energy processes occur already inside the binary 
systems not far away from the massive star. In fact, such modulation of the $\gamma$-ray 
signals from these two binaries (the phases of the maximum emission) have been recently 
predicted based on the calculations of the propagation of high energy $\gamma$-rays 
inside massive binaries (Bednarek~2006a). The optical depths for TeV $\gamma$-rays inside 
these specific binary systems have been also calculated in other papers 
(e.g. B\"otcher \& Dermer~2005, Dubus 2006).

In this paper we concentrate on the application of the detailed Inverse Compton (IC) $e^\pm$ pair cascade model for the binary system LS 5039 (applied already to LS I +61$^{\rm o}$ 303, Bednarek~(2006b, B06)). It is usually expected that such TeV $\gamma$-ray emission can appear in LS 5039 as a result of comptonization of the stellar radiation by relativistic leptons in the 
jet (e.g.  Paredes, Bosch-Ramon \& Romero~2006,  Dermer \& B\"ottcher~2006). Although hadronic production of $\gamma$-rays has been also considered (e.g. Aharonian et al.~2006b, Romero, Christiansen \& Orellana~2005). In our model, we take into account the effects of anisotropic cascade initiated by leptons in the radiation of the massive star including the synchrotron losses. We calculate the $\gamma$-ray light curves and spectra
in the GeV and TeV energy ranges and discuss them in the context of recently observed $\gamma$-ray light curve from LS 5039 (Aharonian et al.~2006a).

The basic parameters of the binary system LS 5039 have been recently
reported by Casares et al.~(2005):
the semi-major axis, $a = 3.4 r_\star$, ellipticity $e = 0.35\pm 0.04$,
the inclination of the binary system toward the observer, 
$i = 24.9\pm 2.8^{\rm o}$ for the case of the black hole with the mass 
3.7 M$\odot$ and $\sim$60$^{\rm o}$ for the neutron star, the azimuthal 
angle of the observer in respect to the periastron passage,
$\omega = 225^{\rm o}$, radius of the massive star, 
$r_\star = 9.3^{+0.7}_{-0.6}$ R$_\odot$, and its surface temperature, 
$T_{\rm s} = 3.9\times 10^4$ K.
For these parameters the distance of the compact object from the 
massive star chances in the range: $r_{\rm p} = 2.2 r_\star$ at the periastron up to
$r_{\rm a} = 4.5 r_\star$ at the apastron.

The binary system LS 5039 shows relativistic radio jets on miliarcsecond scales, with the 
speed of $\sim 0.3c$ (Paredes et al. 2000). It has been identified to be a
counterpart of the EGRET source 3EG J1824-1514 which has relatively flat spectrum 
above 100 MeV with the  spectral index $<2$ (Paredes et al. 2005).
Moreover, the position of LS 5039 is consistent at the $3\sigma$
level with recently detected TeV source HESS J1826-148 (Aharonian et al. 2005). 
The spectrum of this source above 250 GeV is also flat with the photon index $2.12\pm 0.15$,  
although the flux is about two orders of magnitude lower than at GeV
energies. Recent re-analysis of the TeV $\gamma$-ray light curve by Casares et al.~(2005),
using new orbital parameters, has shown possible flux variations of a factor $\sim 3$
with the maximum around the phase $\sim 0.9$. In fact, the HESS Collaboration (Aharonian et al.~2006a) recently detected $\gamma$-ray emission from this source clearly modulated with the orbital period of the binary system with possibly two peaks at the phases $\sim$0.5-0.6 and $\sim$0.8-0.9.

\section{A microquasar in the compact massive binary} 

Details of the IC $e^\pm$ pair cascade model, which we want to apply to LS 5039, have been recently described in Bednarek~(B06, applied for another binary system LS I +61$^{\rm o}$ 303). In this letter, we remind only the most important features of the model, concentrating on the issues which might be characteristic for LS 5039.

\begin{figure}
\vskip 7.5truecm
\includegraphics{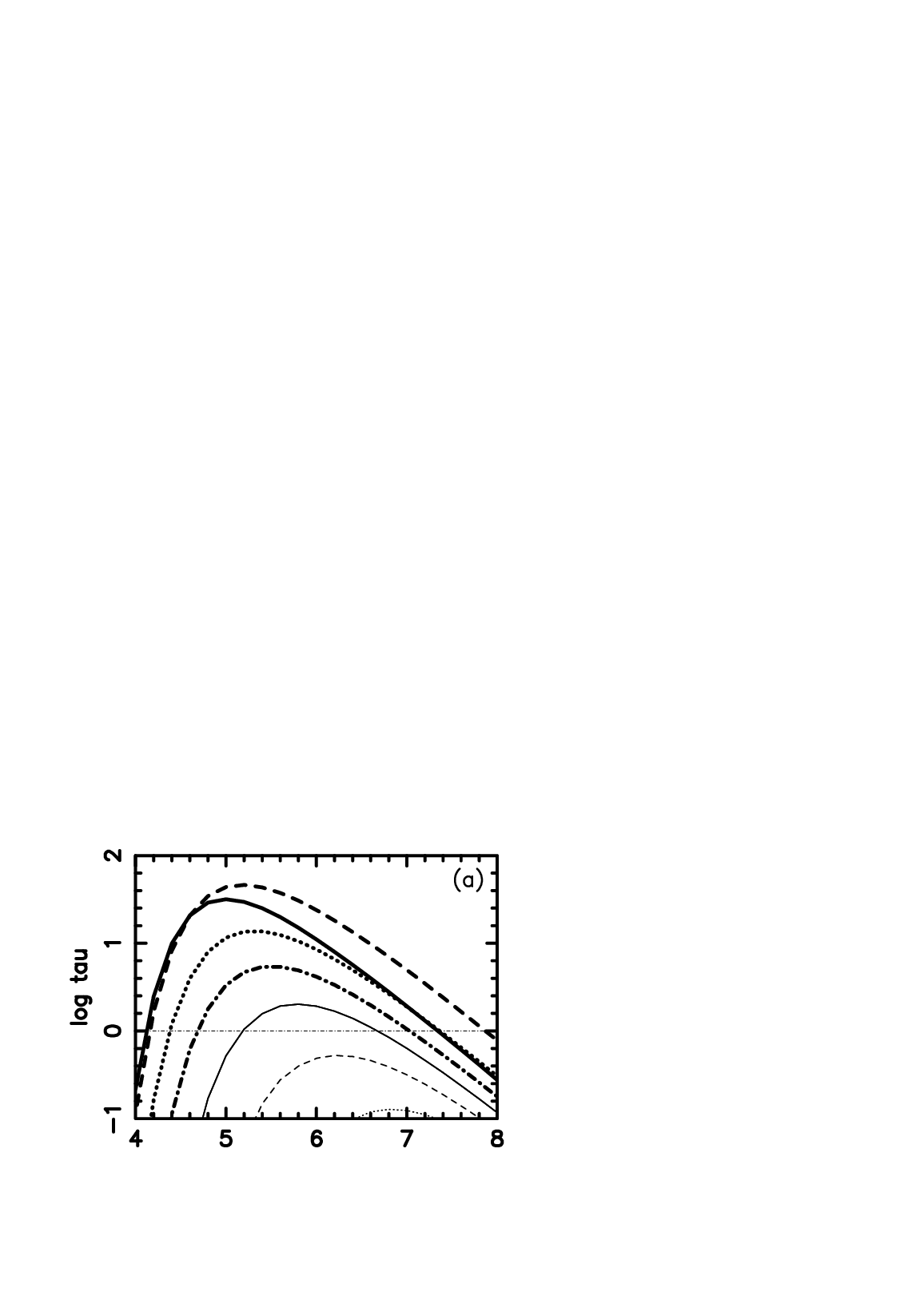}
\includegraphics{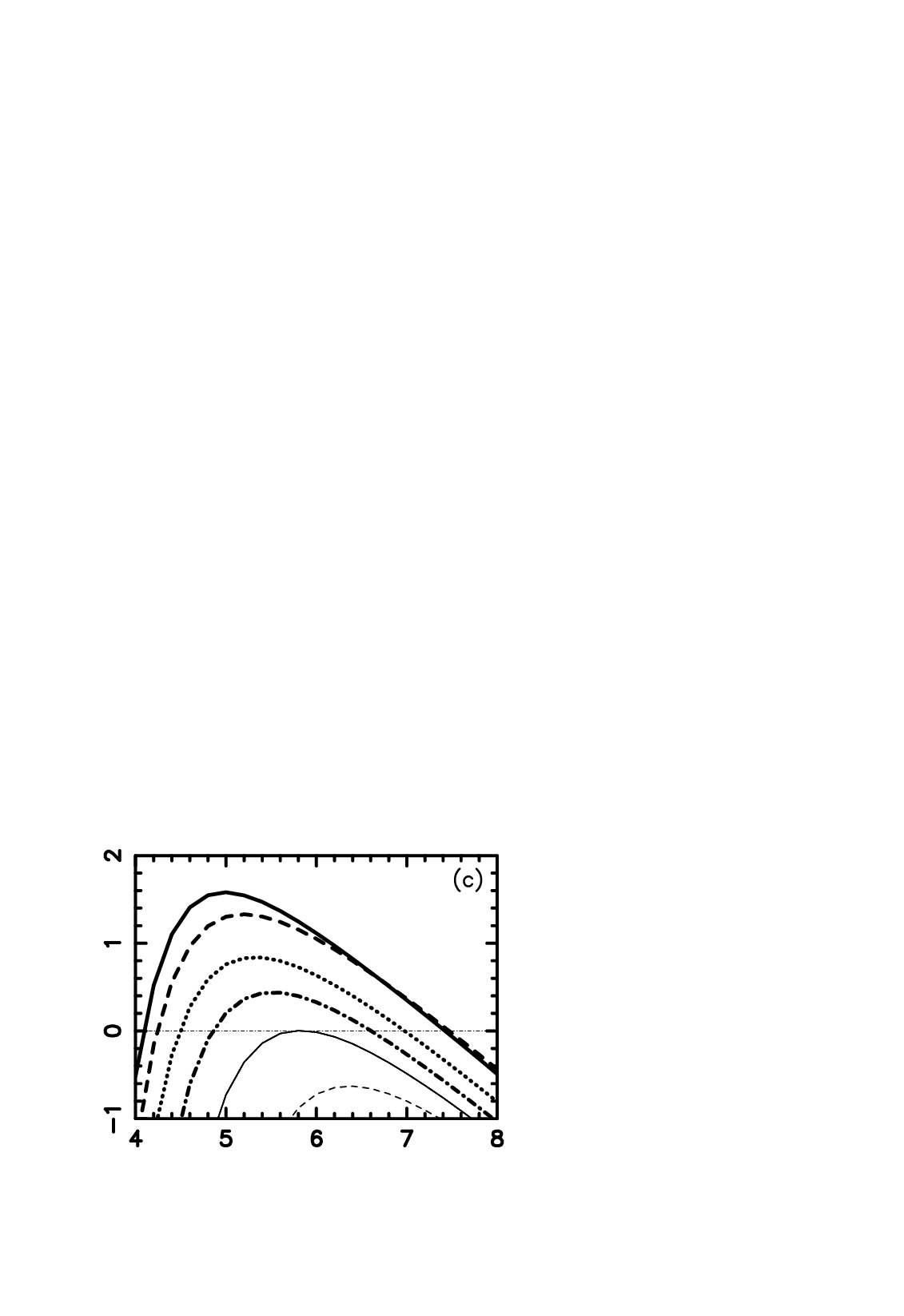} 
\includegraphics{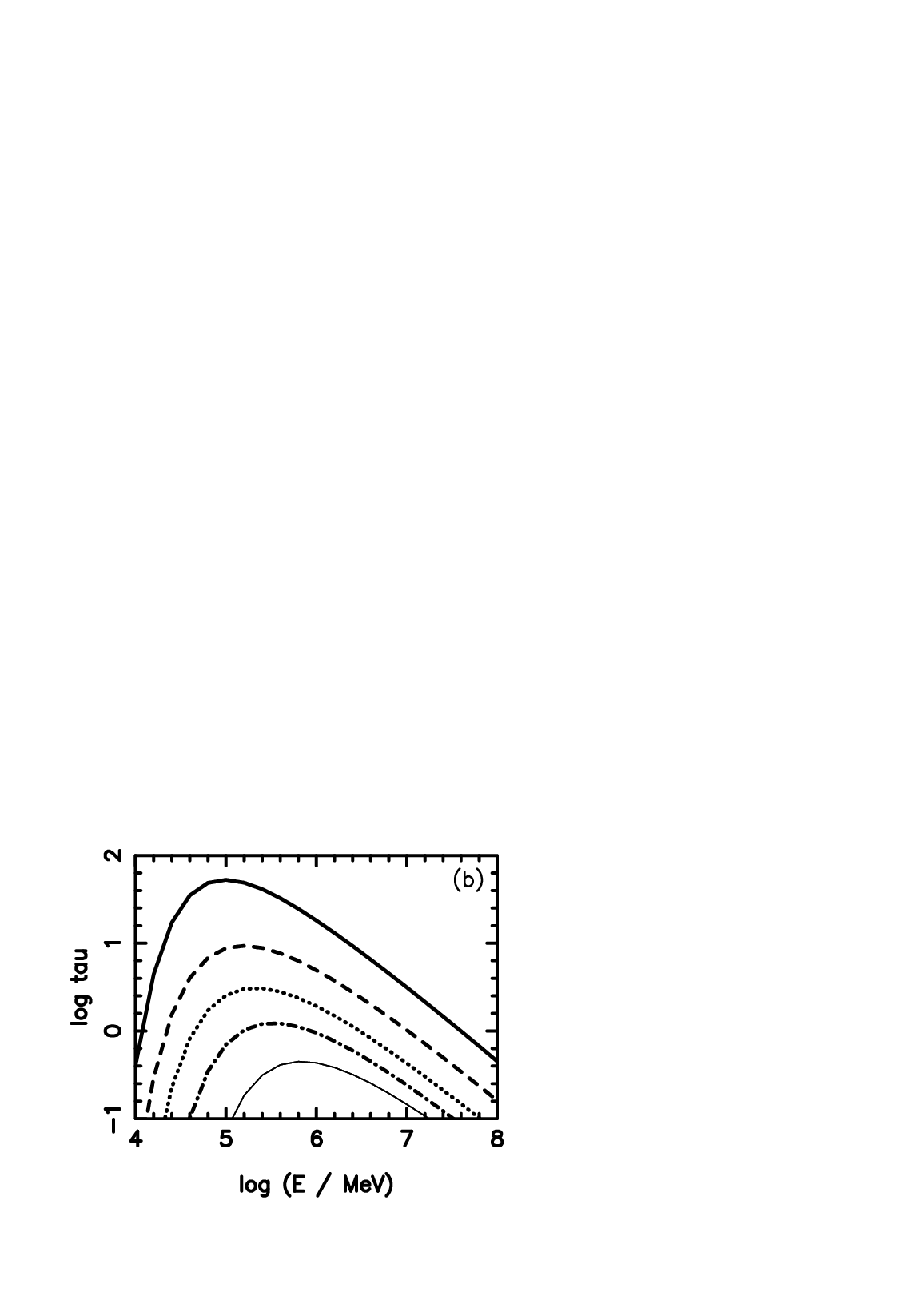}
\includegraphics{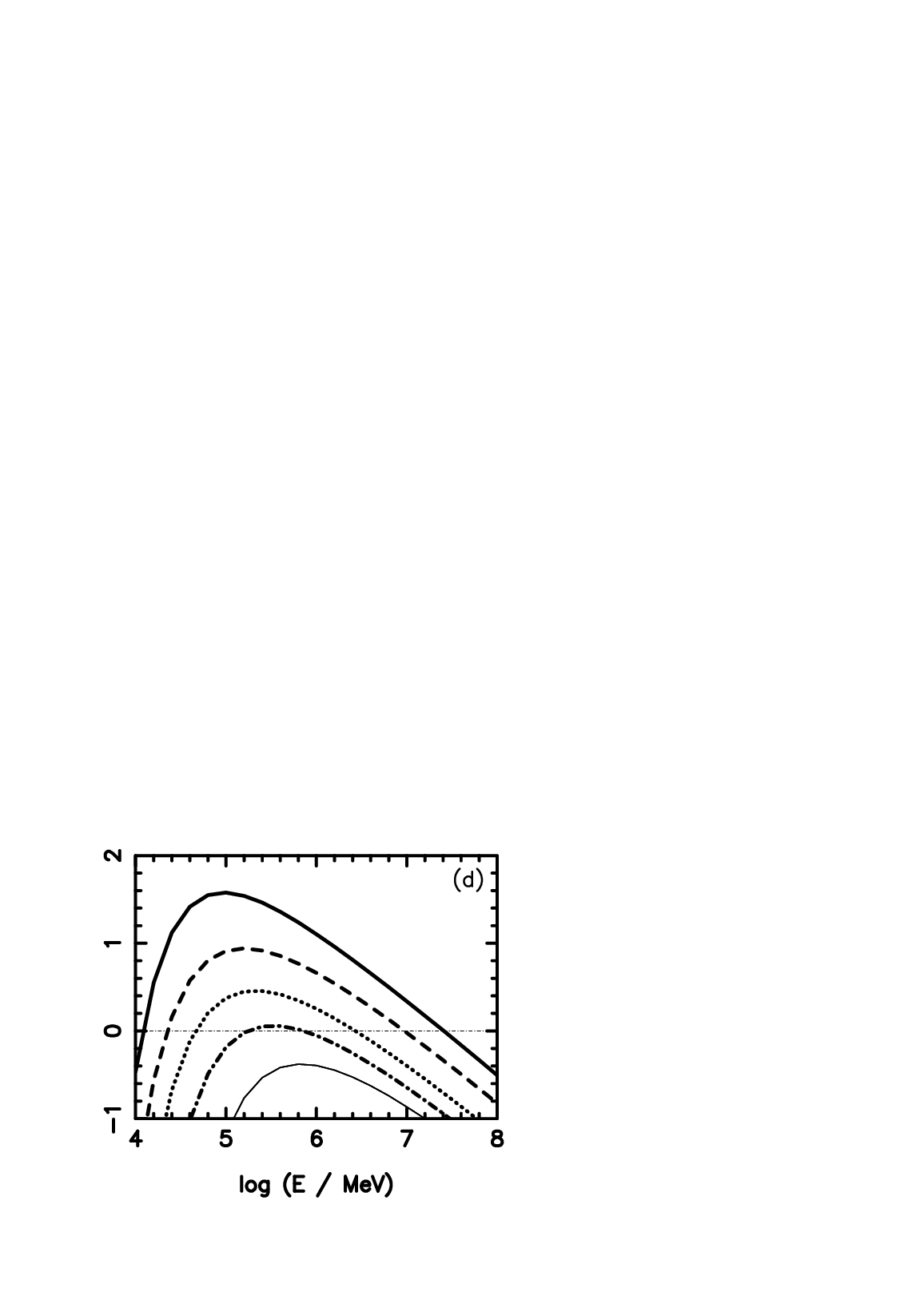}
\caption{The optical depths for $\gamma$-rays (as a function of their energy) 
on the $e^\pm$ pair production in collisions with stellar photons.
$\gamma$-rays are injected in the jet at the distance from its base
$z = 1{\rm r}_\star$ ((a) and (c)),  and $10{\rm r}_\star$
((b) and (d)) for the periastron  passage ($2.2{\rm r}_\star$, left figures) and the apastron 
passage ($4.5 {\rm r}_\star$, right figures) of the compact object on its orbit around the massive star in LS 5039. Specific curves show the optical depths up to the infinity (accept the case in which photons propagate only to the stellar surface) for the injection angles of $\gamma$-rays, $\alpha$,  measured from the direction defined by the centers of the stars, $\alpha = 0^{\rm o}$ (thick full curve, direction toward the 
massive star), $30^{\rm o}$ (thick dashed), $60^{\rm o}$ (thick dotted), 
$90^{\rm o}$ (thick dot-dashed), $120^{\rm o}$ (thin full), 
$150^{\rm o}$ (thin dashed), $180^{\rm o}$ (thin dotted).}
\label{fig1}
\end{figure}

\begin{figure}
\vskip 5.7truecm
\includegraphics{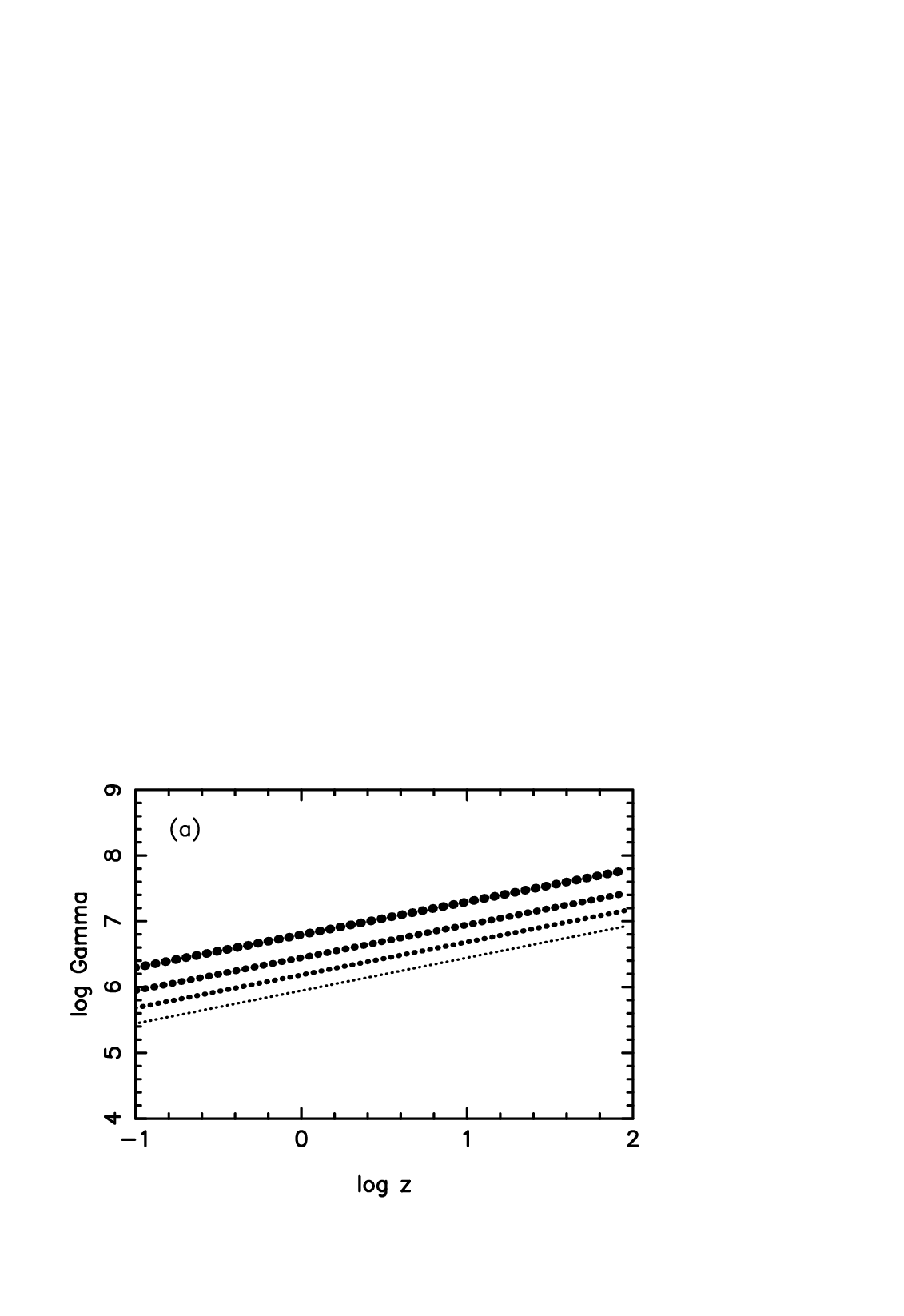}
\includegraphics{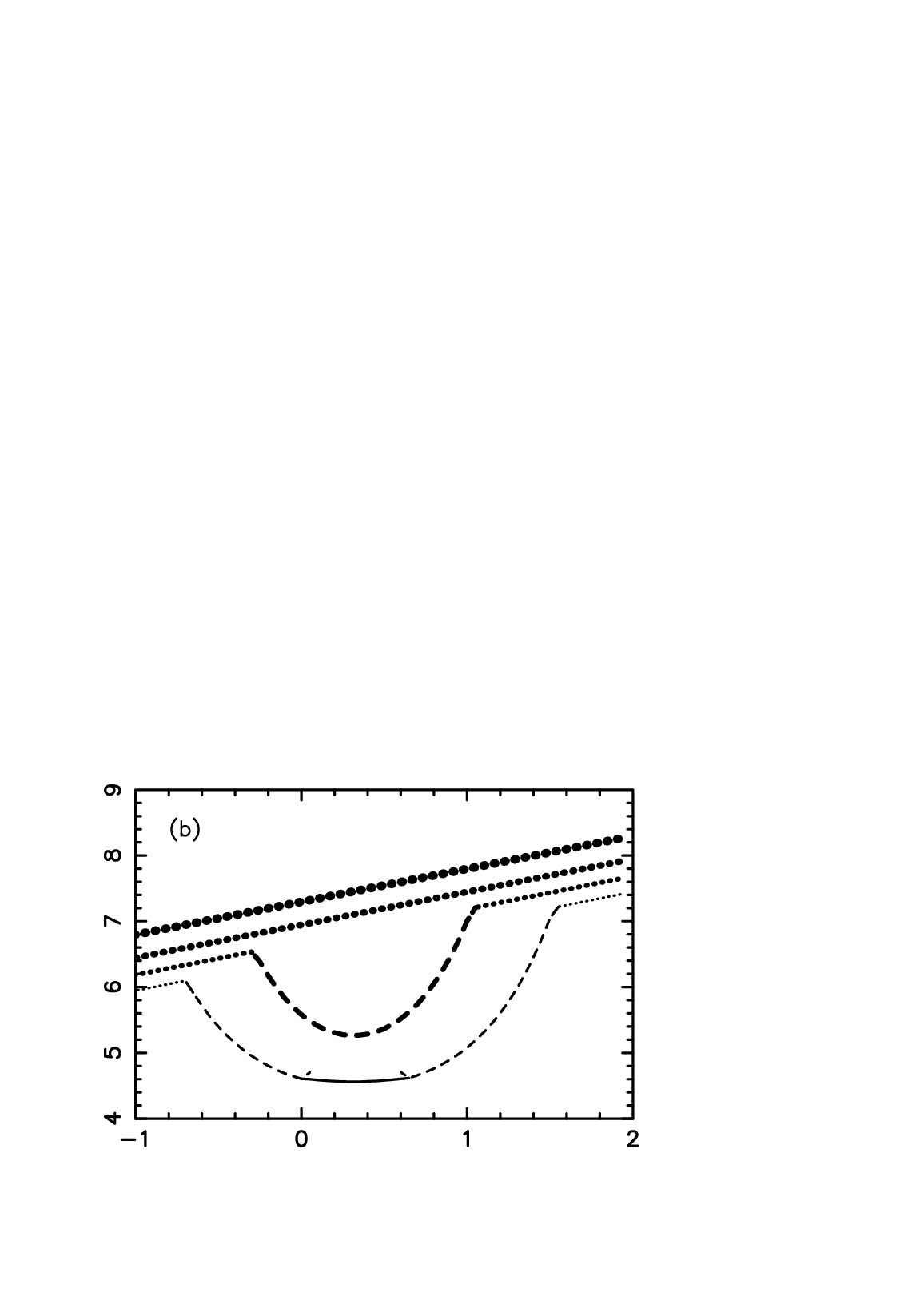}
\includegraphics{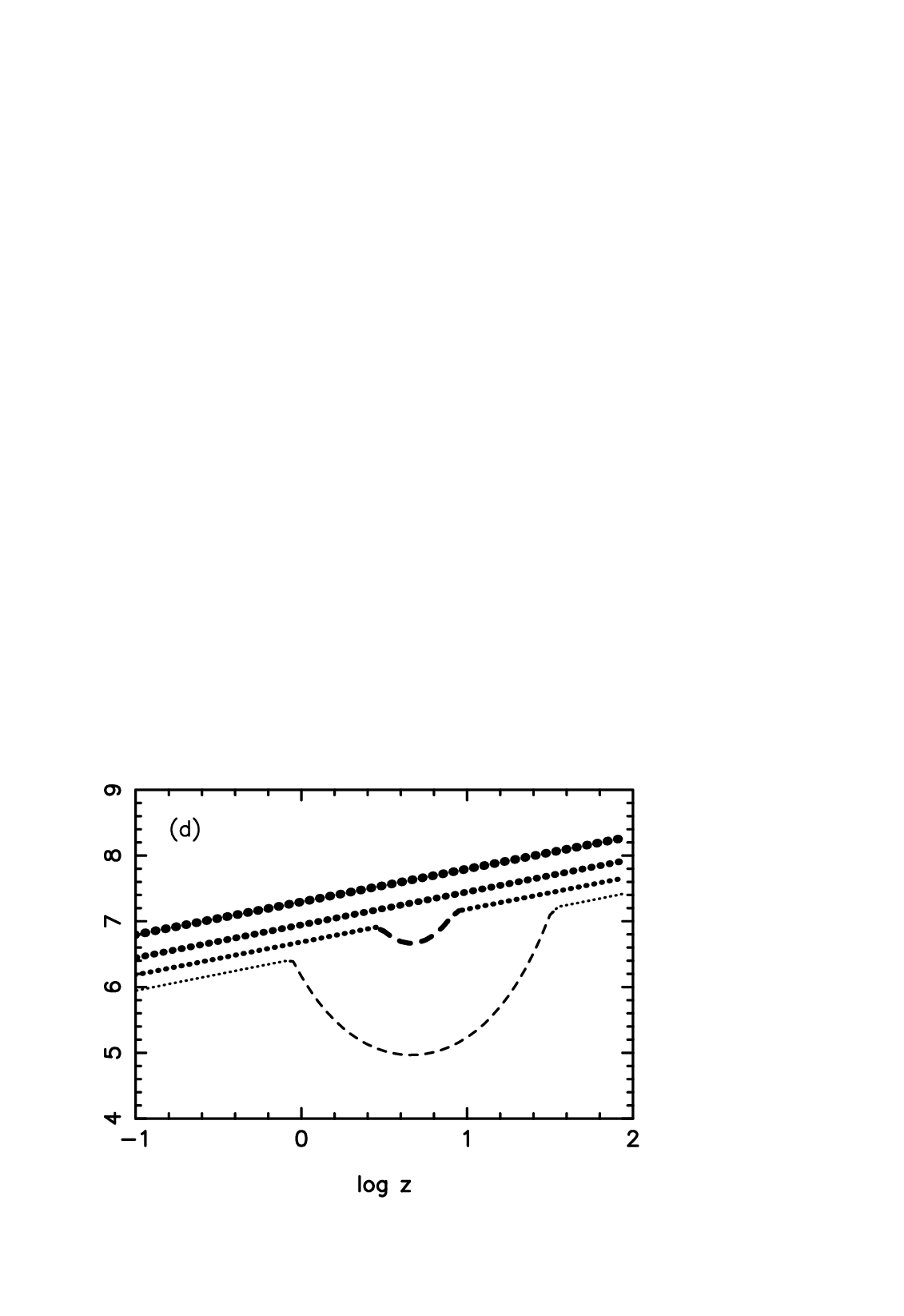}
\includegraphics{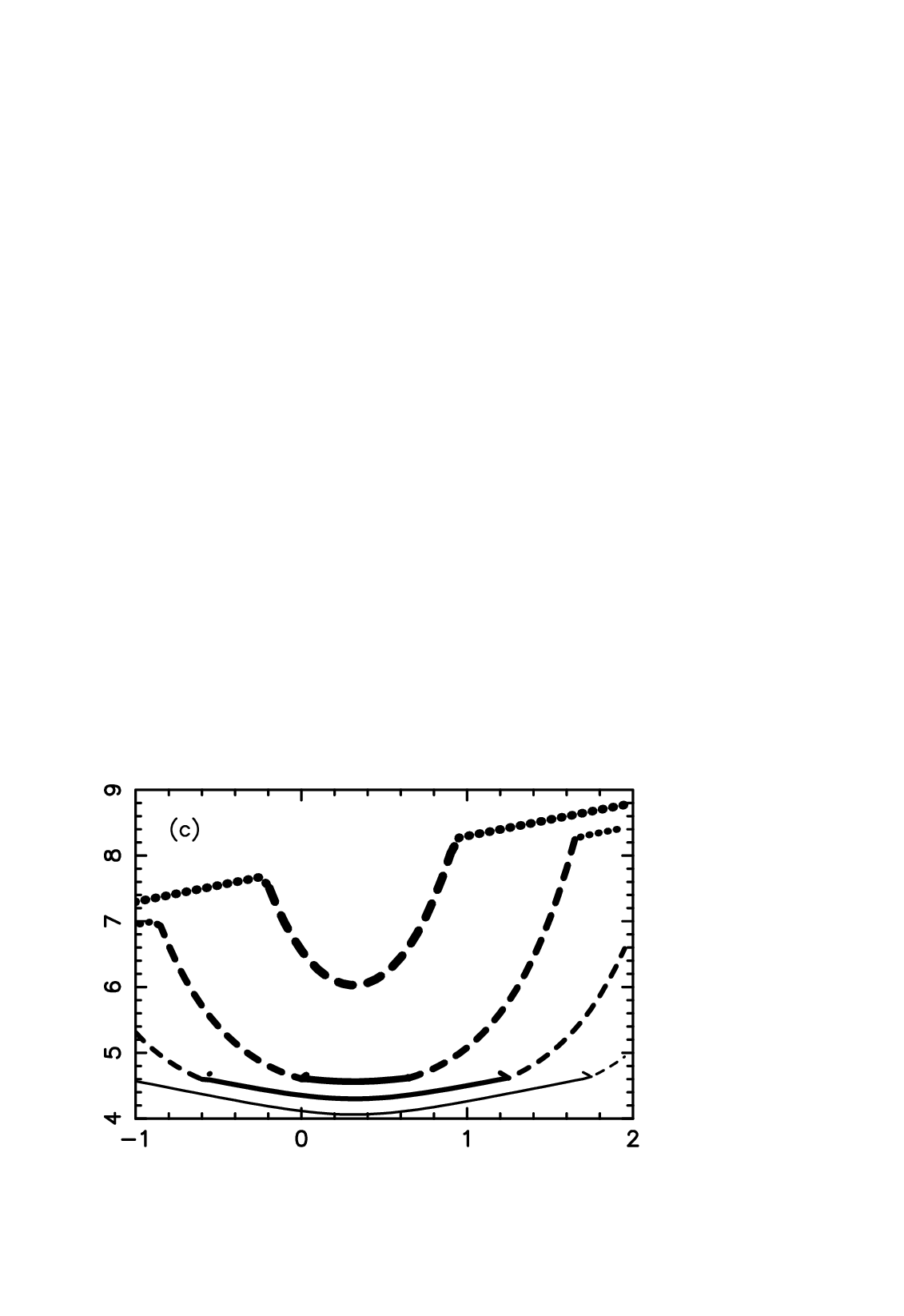}
\includegraphics{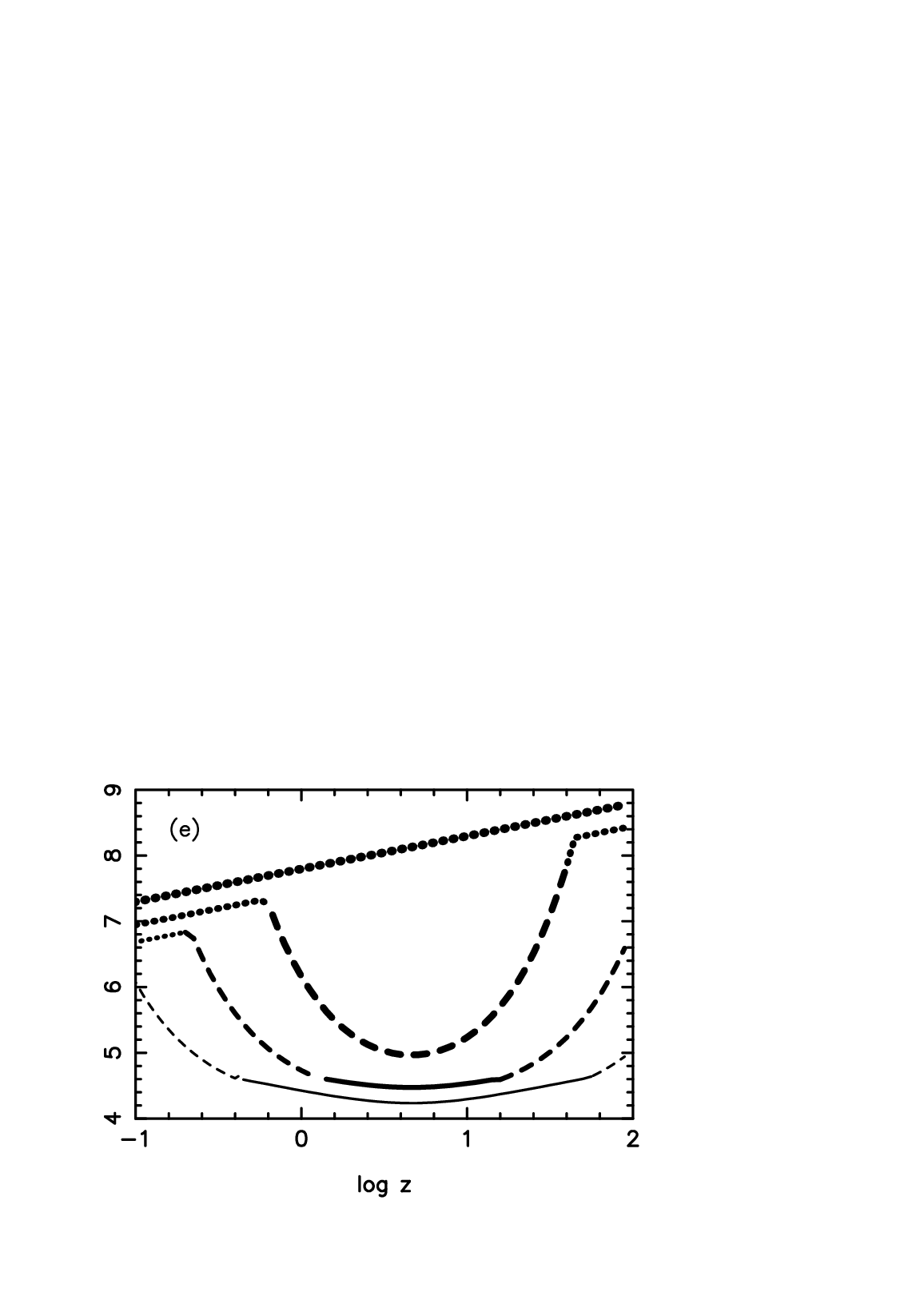}
\caption{The maximum Lorentz factors of electrons as a function of the distance, $z$, 
from the base of the jet (measured in units of the stellar radius) for different 
acceleration efficiencies: $\xi =0.5$ (from the upper, thickest 
curve), 0.1, 0.03, and 0.001 (to the bottom, thinner curve) and 
the parameter describing the magnetic field at the base of the jet  (at at the inner radius of the accretion disk) equal to $B = 3\times 10^5$ G (a)  ($\eta=1$ - the equipartition of magnetic field with disk radiation), $\eta = 0.1$ ((b) and (d)), and 0.01 ((c) and (e)). The Maximum Lorentz factors are obtained from the comparison of the acceleration rate with the energy loss rate on synchrotron process (dotted lines), inverse Compton process in the Thomson regime (full curve) and the Klein-Nishina regime (dashed curves), for the periastron 
(figures
(a), (b), and (c)) and the apastron ((a), (d), and (e)) passages of the compact object.}
\label{fig2}
\end{figure}

Leptons injected into the radiation field of the massive star initiate the IC $e^\pm$ pair cascade provided that, at least in some directions, the optical depths for $\gamma$-rays 
(in $\gamma + \gamma\rightarrow e^+e^-$ process) are above unity.
We calculate such optical depths in the general case, i.e. for an arbitrary place of injection of $\gamma$-ray photons with arbitrary energies and angles of propagation,
applying the parameters of the massive star in LS 5039 (see Fig.~\ref{fig1}).
For certain range of photon energies and their injection angles,
the optical depths are clearly above unity. Thus, the basic conditions for the application of such a model are fulfilled since the electrons in the jet preferentially produce the first generation $\gamma$-rays in directions of the largest optical depth. We consider a simple geometrical scenario in which two sided jet is launched along the disk axis. For simplicity, we assume that 
the surface of the disk is in the plane of the binary system, i.e. the jet direction 
is perpendicular to the plane of the binary system. Relativistic leptons are 
injected along the jet at some range of distances from its base, starting at $z_{\rm min}$ up to $z_{\rm max}$. The magnetic field strength, which determine
the process of acceleration of leptons in the jet and their losses on synchrotron process,
is described by a parameter $\eta$ (the ratio of the magnetic field energy density to radiation energy density in the inner part of the accretion disk, see details of the model for the magnetic field in the jet in Bednarek~(B06)). The energy losses of leptons on IC scattering in the Thomson and the Klein-Nishina regime are important as well. They are taken into account when considering the acceleration of electrons. We assume that leptons are accelerated with a power law spectrum (and spectral index which does not depend on $z$) in multiple shocks propagating along the jet. 
The maximum energies of electrons are determined by the acceleration mechanism (defined by the acceleration parameter $\xi$) and the energy losses
on the most efficient radiation process (see B06). These maximum energies are shown for specific values of $\xi$ and $\eta$ as a function of the distance, $z$, from the base of the jet (Fig.~\ref{fig2}). It is clear that for reasonable values, leptons can reach energies over a few tens of TeV, already at distances from the base of the jet closer than 10 radii of the massive star. We consider two different scenarios for the injection rate of relativistic leptons along the jet. The injection rate (electrons per unit length along the jet per unit time) is
constant along the jet from its base at $z_{\rm min}$ up to $z_{\rm max}$ (i.e. $N(z)\propto const$), or it drops along the jet proportionally to the magnetic energy density in the jet starting from some distance, $z_{\rm min}$, from the base of the jet (i.e. $N(z)\propto z^{-2}$).

\section{The cascade $\gamma$-ray spectra}

\begin{figure}
\vskip 6.5truecm
\includegraphics{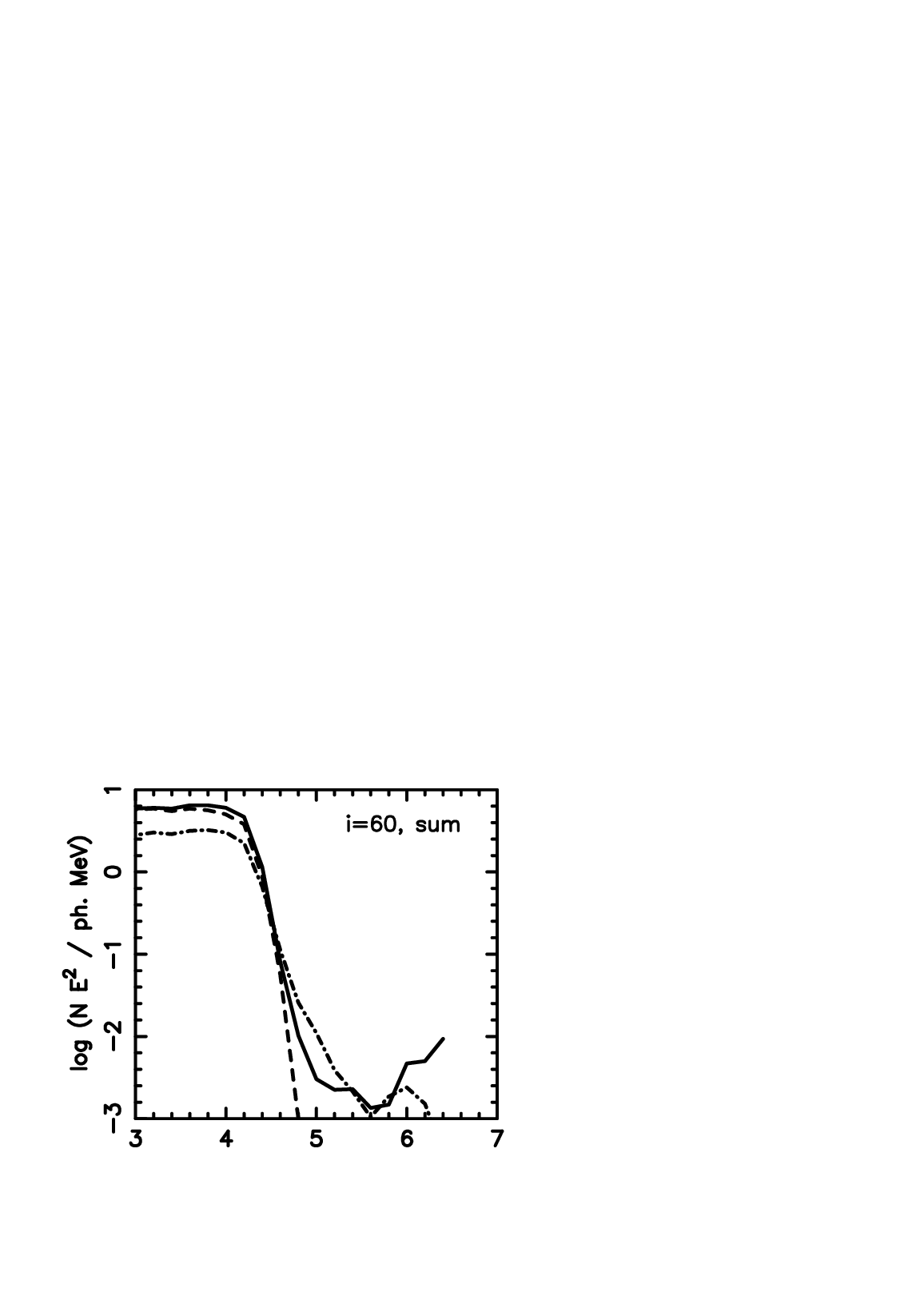}
\includegraphics{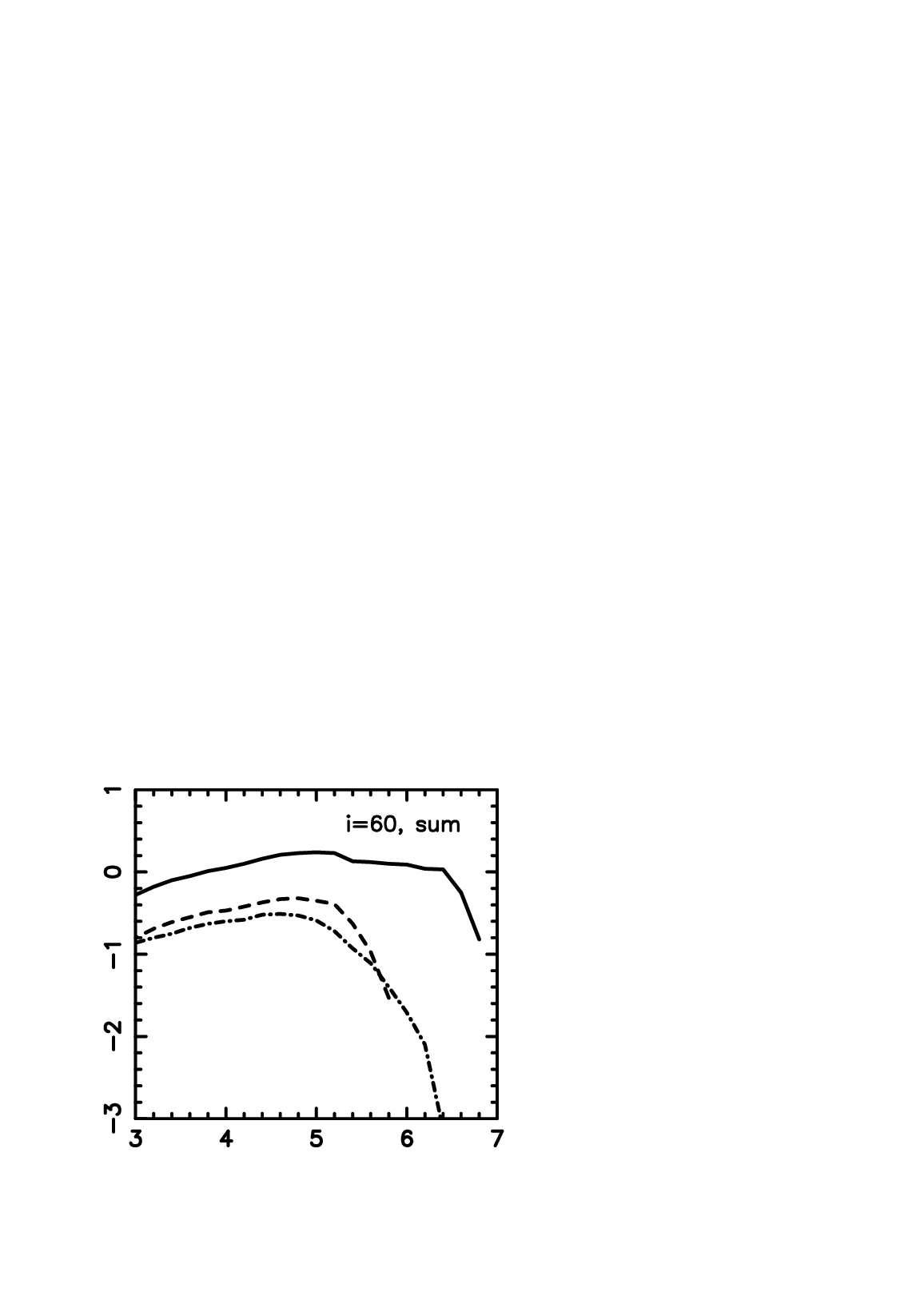}
\includegraphics{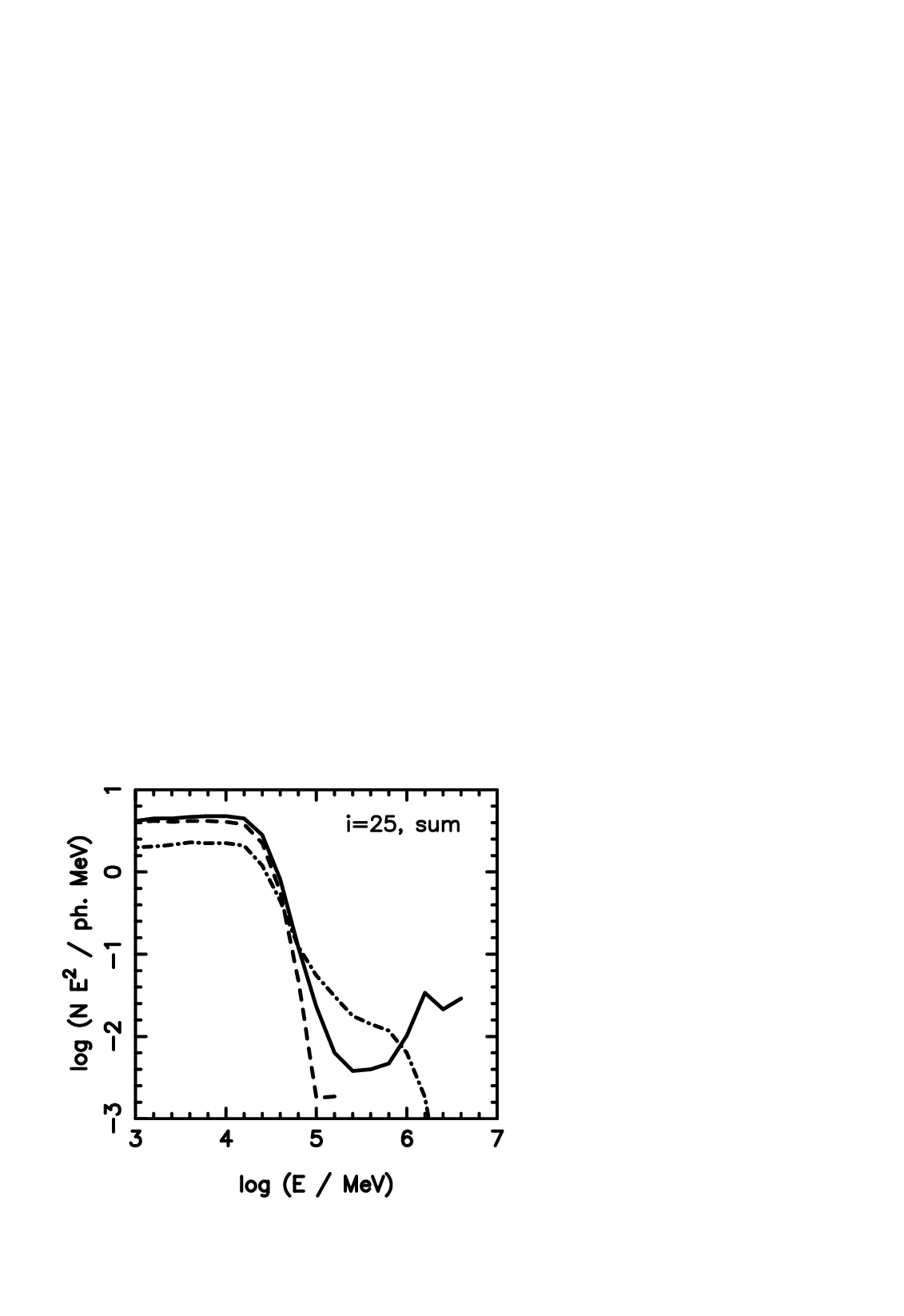}
\includegraphics{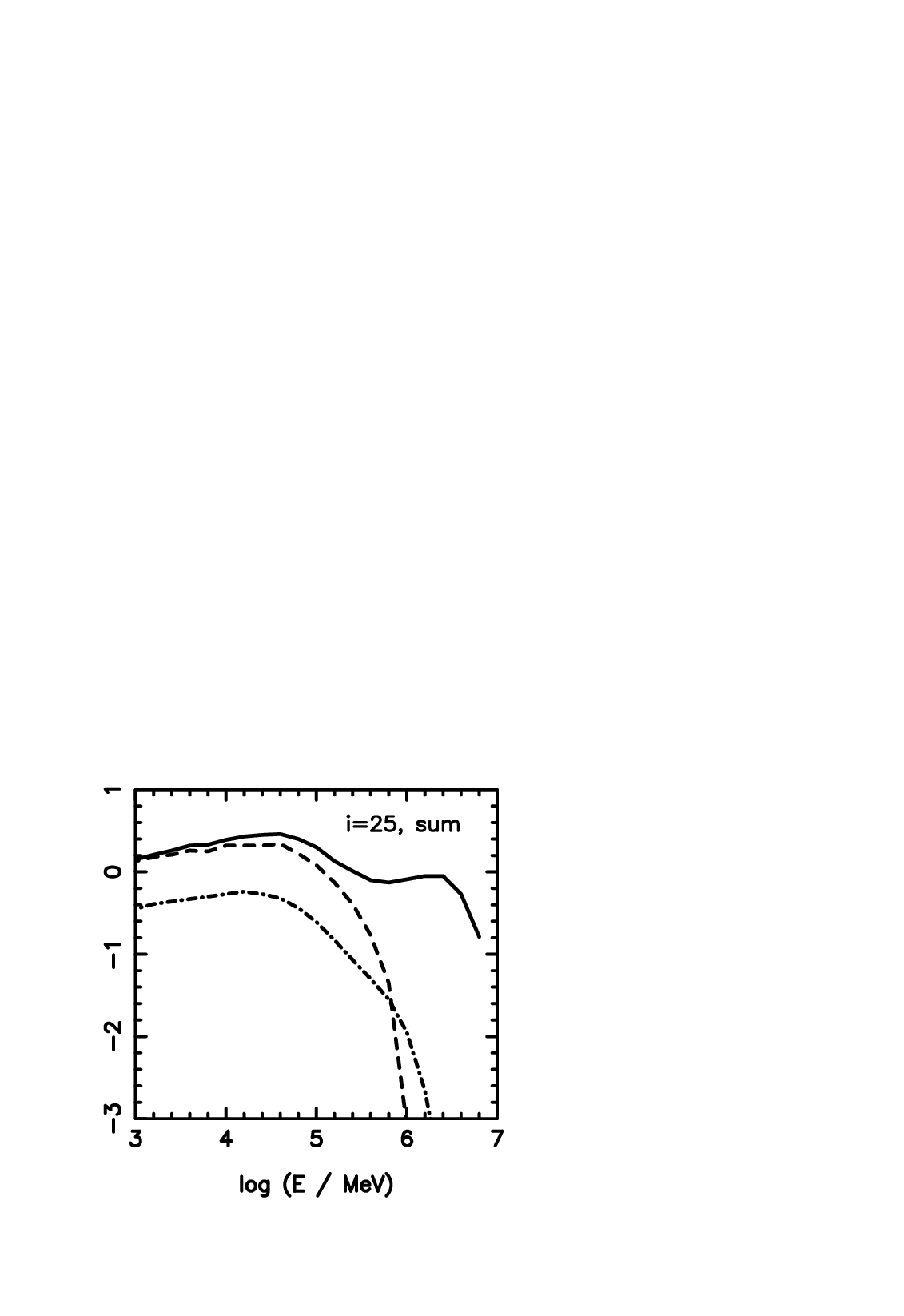}
\caption{Spectral Energy Distributions (SED) 
from cascades initiated by primary electrons injected in the jet with: the efficiency of electron injection depending on the distance from the base of the jet as $N(z)\propto z^{-2}$ (independently on 
the phase of the binary system), and differential power law spectrum of electrons with the index equal to $-2$. The left figures show the $\gamma$-ray spectra, produced at 
the periastron passage of the compact object for the observer located at the inclination angles
$i = 25^{\rm o}$ and $60^{\rm o}$, from the {\it jet} and  {\it counter-jet} (the {\it sum} of both) . The jets propagate perpendicular to the plane of the binary system. The $\gamma$-ray spectra produced at the apastron
passage of the compact object are shown on the right figures.  
The specific spectra are calculated for the acceleration conditions in the jet described by 
the parameters: $\xi =0.03$ and $\eta=0.1$  (full curves),
$\xi =0.3$ and $\eta=0.01$, (dashed), and $\xi =0.3$ and $\eta=0.1$ 
(dot-dashed).
}
\label{fig3}
\end{figure}

Since leptons are injected isotropically in the jet, they produce $\gamma$-rays preferentially in directions where they find the largest optical depth.
Therefore, absorption of $\gamma$-rays and interaction of secondary $e^\pm$ pairs have to be taken into account when calculating the $\gamma$-ray spectra produced
by leptons in the radiation field of the massive star both at the periastron and the apastron passages. We have performed  the Monte Carlo calculations of the $\gamma$-ray spectra escaping from the binary system at arbitrary
directions, applying the above mentioned model for the acceleration of primary leptons
in the jet.
To allow a simple analysis, at first we consider the case of injection of leptons in the jet
with the power law spectrum, and differential spectral index equal to $-2$, with  
the cut-off which depends on the location of the acceleration place in the jet and on the phase of the compact object (see Fig.~2). These cut-offs are determined by the values of $\eta$ and $\xi$.
Since two jets are expected from a single compact object, we  consider the production of $\gamma$-rays in the jet which propagate above the plain of 
the binary system (i.e. the {\it jet} directed towards the hemisphere containing the observer) and in the jet propagating below the plane of the binary system 
(i.e. the {\it counter-jet}), see (B06). 
The results of example calculations of the $\gamma$-ray spectra from the IC $e^\pm$ pair cascade are shown in Fig.~\ref{fig3} for different values of $\xi$ and $\eta$. The observer is located at the inclination angles of $25^{\rm o}$ and $60^{\rm o}$. The spectra are shown for  the compact object at the periastron (left figures) and the apastron passages (right). 
The $\gamma$-ray spectra, produced by leptons injected in the jet at the periastron passage, steepen significantly at energies above a few tens of GeV due to the efficient cascading. This cascading effects are significantly smaller when the compact object is close to the apastron. 
However, note that the $\gamma$-ray spectra calculated for these parameters do not fit nicely to the spectra observed recently by Aharonian et al.~(2006a) due to the deficit of $\gamma$-ray emission above $\sim 100$ GeV. Therefore, in the next section we 
consider the model with the flatter profile for the injection of electrons into the jet
(i.e. $N(z) = const$).

\section{Confrontation with observations of LS 5039}

\begin{figure}
\vskip 7.5truecm
\includegraphics{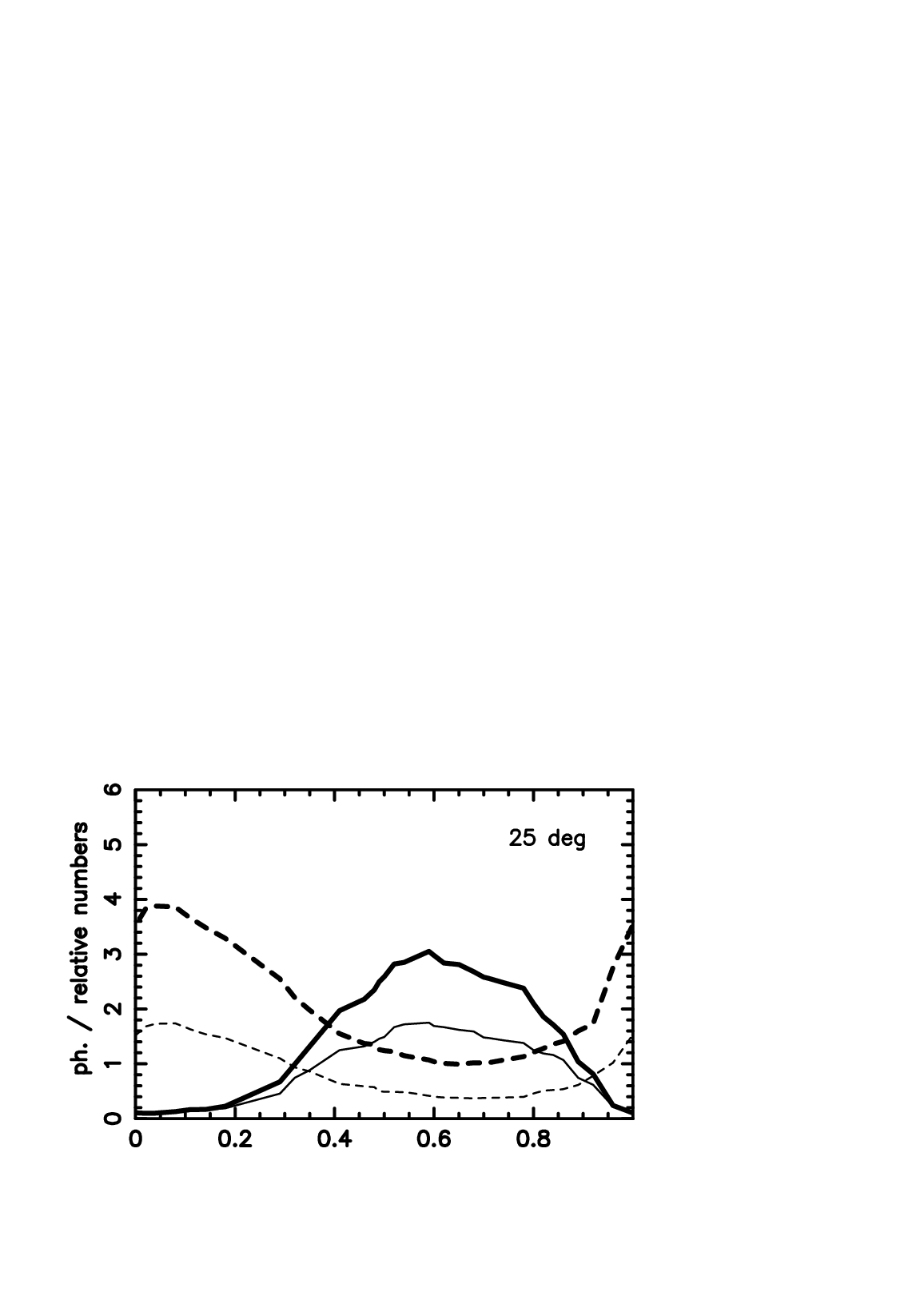}
\includegraphics{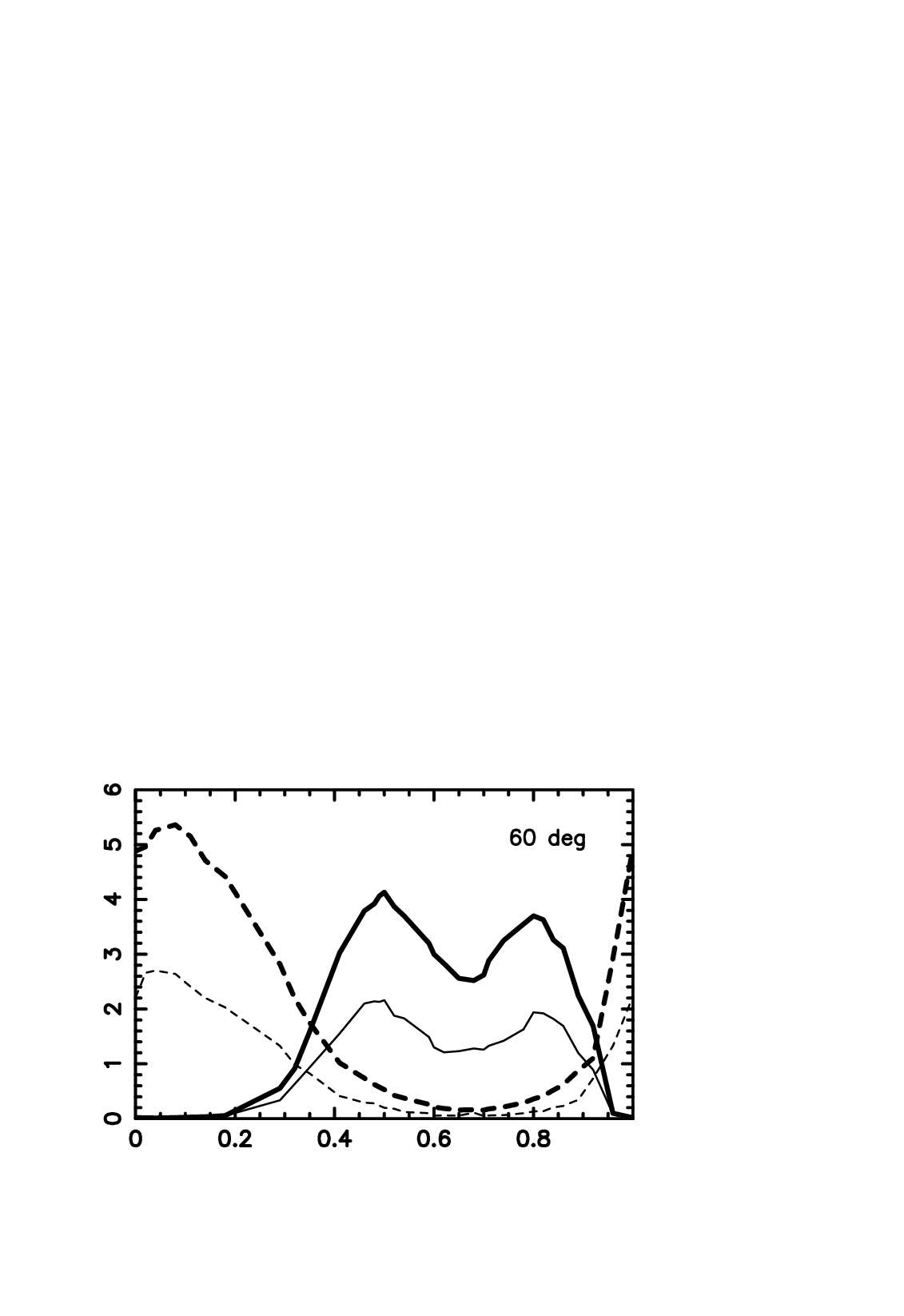}
\includegraphics{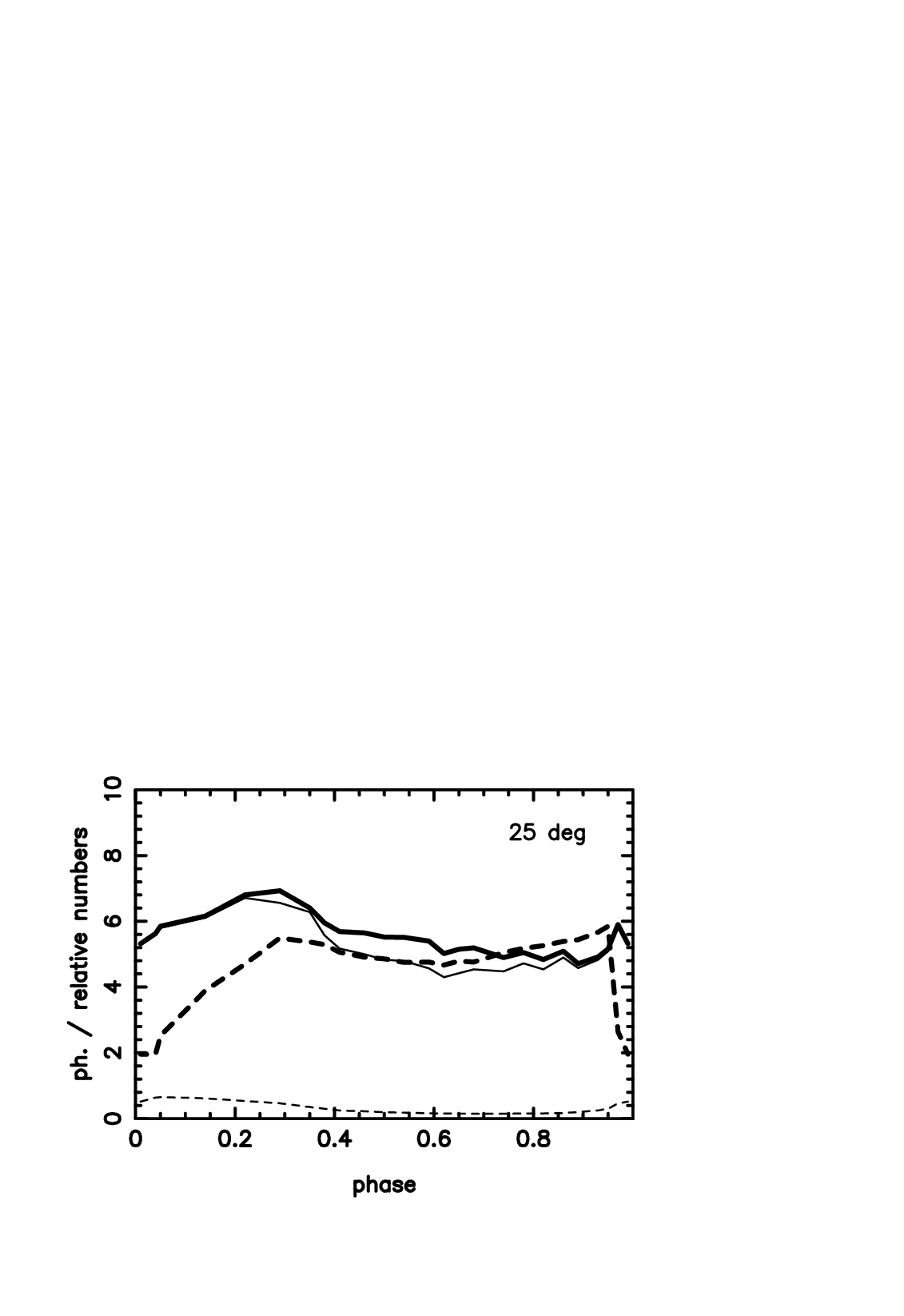}
\includegraphics{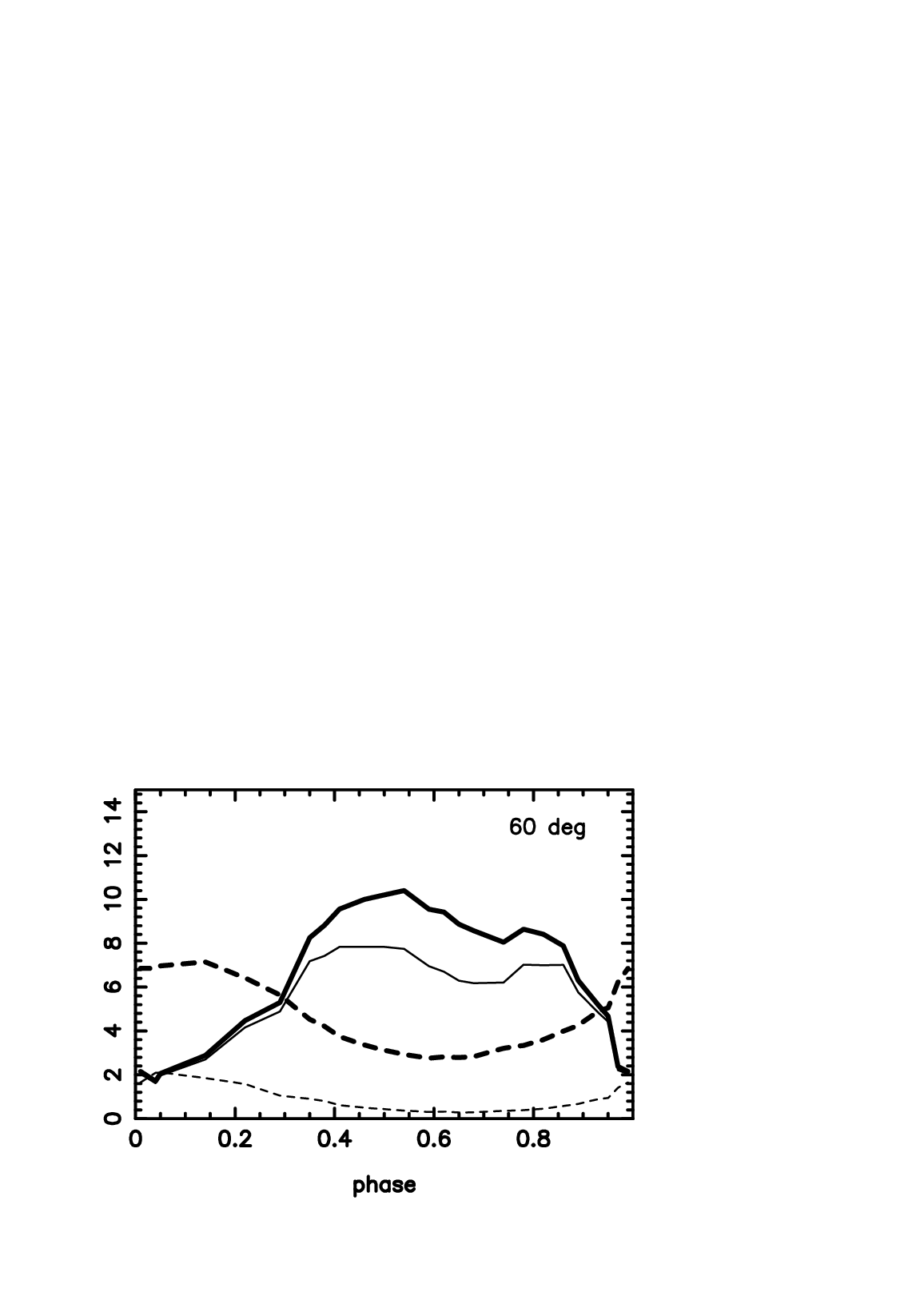}
\caption{The $\gamma$-ray light curves in the energy range $1-10$ GeV (dashed curves) and 
$>$300 GeV (full curves), produced by electrons in the jet (thin curves) and the jet + counter-jet (thick curves), are shown on the upper panel for two possible inclination angles of the binary system LS 5039 ($25^{\rm o}$ and $60^{\rm o}$), $\xi = 0.1$, $\eta= 0.1$, $z_{\rm min} = 0.1r_\star$, and $z_{\rm max} = 10r_\star$. On the bottom panel, the $\gamma$-ray light curves are shown for $\xi = 0.2$, $\eta= 0.1$, and $z_{\rm max} = 20r_\star$.}
\label{fig4}
\end{figure}

Let us select the set of parameters which allow acceleration of electrons to energies
up to $\sim$10 TeV (see upper panels in Fig.~\ref{fig2}). For these parameters 
($\xi = 0.1$ and $\eta = 0.1$) and the case of electrons injected with the constant rate 
along the jet between $z_{\rm min} = 0.1r_\star$ and $z_{\rm max} = 10r_\star$, we calculate the 
$\gamma$-ray light curves in two energy ranges, i.e 1-10 GeV (the GeV region, 
the AGILE and GLAST energy range) and $>$300 GeV (the TeV region, the Cherenkov telescopes 
energy range), and two possible inclination angles of the binary system LS 5039, i.e. 
$25^{\rm o}$ and $60^{\rm o}$ (the range of angles derived by Casares et al.~2005). 
The TeV $\gamma$-ray light curves, calculated for $25^{\rm o}$, show broad maximum between 
the phases $\sim 0.4-0.8$ peaked 
at the phase $\sim 0.6$. On the other hand, the light curves for the inclination angle $60^{\rm o}$ show more 
complicated structure. The broad maximum between the phases $\sim 0.4-0.8$ shows clear two additional
peaks centered at the phases $\sim$0.5 and $\sim$0.8 and a dip at the phase $\sim 0.65$
(Fig.~\ref{fig4}). 
The general broad maximum observed for both inclination angles is consistent with 
recently reported TeV $\gamma$-ray light curve observed by the HESS Collaboration 
(see the bottom panel in Fig.~5 in Aharonian et al.~2006a). Note however, that the HESS 
TeV $\gamma$-ray light  curve also shows,
although not statistically very significant, a dip in the maximum at the phase $\sim 0.65$, i.e. corresponding to the interior conjunction of the compact object.
The confirmation of such a structure by the future HESS observations will strongly argue
for the inclination of the binary system closer to $60^{\rm o}$, supporting the hypothesis
on the presence of a neutron star in this binary system.

Note, that recently B\"ottcher~(2006) argues for the inclination angle of the LS 5039 
closer to $30^{\rm o}$, based on the deabsorption of the $\gamma$-ray spectrum 
observed by the HESS experiment from LS 5039. However, these calculations base on the assumption that the cascading processes initiated by secondary leptons (originated in the absorption process of TeV $\gamma$-rays) in the radiation of the massive star can be neglected.
This is valid provided that the magnetic field inside the volume of the binary system is
strong enough to guarantee efficient cooling of secondary pairs on synchrotron process
(in the case of LS 5039 should be of the order of $\sim 100$ G).
In another case, as discussed in this paper, the IC $e^\pm$ pair cascades should be 
taken into account since the cross sections for $\gamma$-$\gamma$ absorption and IC scattering in the KN regime are similar, see e.g. the optical depths for these two processes in Fig.~2 in Sierpowska \& Bednarek~(2005).

In our model the broad maximum between phases $\sim 0.4-0.8$ in the TeV $\gamma$-ray light 
curve is due to the favorite location of the compact object in respect to the massive star and the observer.
At these phases the jet is generally on the other site of the massive star in respect to the observer and the conditions for scattering of stellar
radiation by electrons in the jet into the TeV $\gamma$-ray energy range are optimal. 
However, when the line of sight to the compact 
object passes very close to the massive star, which occurs in the case of large inclination angles,
absorption of TeV $\gamma$-rays in the stellar radiation start to play dominant role.
This is the reason of the appearance of additional dip in the TeV light curve at phase $\sim 
0.65$ for large inclination angles of the binary system. For low inclination angles,
the line of sight is sufficiently away from the massive star and the absorption effects of TeV $\gamma$-rays are too low to produce additional dip in the light curve.

On the other hand, the GeV $\gamma$-ray light curve behaves very different. 
It is clearly anti-correlated with the TeV $\gamma$-ray emission. This is caused by more 
efficient cascading 
close to the periastron passage of the compact object in which many GeV photons are created. 
In directions where the optical depths of TeV $\gamma$-rays are large, the 
energy from electrons, originally transfered to primary TeV $\gamma$-rays, is degraded to
the GeV $\gamma$-rays. In contrary, in directions of low optical depths, TeV $\gamma$-rays
escape with relatively mild absorption producing the peak in the TeV light curve.
Therefore, the maximum in the GeV
$\gamma$-ray light curve appears just after periastron, at the phase $\sim$0.05 
(i.e. at the superior conjunction of the compact object), and the minimum at the phase 
$\sim$0.7 (i.e. at the inferior conjunction). It is expected that the GeV $\gamma$-ray emission 
may vary by a factor up to $\sim$30 with the period of  LS 5039 (see Fig.~\ref{fig4}). 
Hence, we predict also strong modulation of the GeV emission from LS 5039 with its orbital 
period which should be easily observed by the AGILE and GLAST detectors.

Note that the TeV $\gamma$-ray light curve, reported by Aharonian et al.~(2006a, Fig~5), show clear base line emission also at the periastron passage of the compact object. 
In order to check if this emission can originate farther along the jet, we show in the bottom panel in Fig.~\ref{fig4} the $\gamma$-ray light curves for the case of injection of electrons
in the jet up to  $z_{\rm max} = 20r_\star$,  $\xi = 0.2$, and $\eta = 0.1$.  In fact, such base line emission at a level of $\sim$20$\%$ of the peak emission appears close to the periastron for the inclination angle $i = 60^{\rm o}$. 
Moreover, previously found two peak structure in the TeV $\gamma$-ray light curve (or 
broad high level emission through the phases $\sim$0.4-0.8) for the case of 
inclination angle of the binary system $i = 60^{\rm o}$ is still present.
However, in contrary to the previously discussed case in the upper panels in Fig.~4, the TeV $\gamma$-ray light curve for the inclination angle $25^{\rm o}$ is quite flat due to similar conditions farther along the jet  independently on the periastron or apastron location of the compact object.

The spectral features observed by the HESS (Aharonian et al.~2006a), flat power spectrum at the inferior conjunction and steep spectrum at the superior conjunction, can be also explained in terms of the considered here general model. Such features appear even in so simplified model for  the injection of relativistic electrons in the jet  (see Fig.~3), i.e single power law spectra for the electrons and 
their injection rate independent on the phase of the compact object. Based on these results we conclude that a more realistic model  (which relax these simple assumptions) can
certainly produce exact fitting of the experimental spectral data, although huge
computation time requirements prevent such direct comparison at present.
Since in the present model the cooling of electrons in the jet on IC process is very efficient
(Bednarek~2006b), and the injection rate of electrons does not depend on the phase of the compact object, the total requirement on the energy budget of accelerated electrons is comparable to the power emitted in the $\gamma$-rays at the high state, i.e. close to the inferior conjunction. 

\begin{acknowledgements}
I would like to thank the anonymous referee for valuable comments.
This work is supported by the Polish MNiI grant No. 1P03D01028.
\end{acknowledgements}


\begin{thebibliography}{}

\bibitem{} Aharonian, F. et al. (HESS collab.) 2005 Sci. 309, 746
\bibitem{} Aharonian, F. et al. (HESS collab.) 2006a A\&A, in press (astro-ph/0607192)
\bibitem{} Aharonian, F., Anchordoqui, L., Khangulyan, D., Montaruli, T. 2006b J.Phys.Conf.Ser. 39, 408
\bibitem{} Albert, J. et al. (MAGIC collab.) 2006 Sci. 312, 1771
\bibitem{} Bednarek, W. 2006a, MNRAS 368, 579
\bibitem{} Bednarek, W. 2006b, MNRAS 371, 1737 (B06) 
\bibitem{} B\"ottcher, M. 2006, APh, submitted (astro-ph/0609136)
\bibitem{} B\"ottcher, M., Dermer, C.D. 2005, ApJ 634, L81 
\bibitem{} Casares, J., Ribo, M., Ribas, I., Paredes, J.M., Marti, J., Herrero, A. 2005 
MNRAS 364, 899
\bibitem{} Dermer, C.D., B\"ottcher, M. 2006, ApJ 643, 1081
\bibitem{} Dubus, G. 2006 A\&A 451, 9
\bibitem{} Paredes, J.M., Bosch-Ramon, V., Romero, G.E. 2006, A\&A 451, 259
\bibitem{} Paredes, J.M., Marti, J., Ribo, M., Massi, M., 2000 Sci. 288, 2340
\bibitem{} Romero, G.E., Christiansen, H.R., Orellana, M. 2005 ApJ 632, 1093
\bibitem{} Sierpowska, A., Bednarek, W. 2005, MNRAS 356, 711 

\end{thebibliography}
\end{document}